\documentclass[english]{article}
\usepackage[T1]{fontenc}
\usepackage[latin9]{inputenc}
\usepackage{verbatim}
\usepackage{textcomp}
\usepackage{amstext}
\usepackage{graphicx}
\usepackage{amssymb}

\makeatletter

\newcommand{\lyxmathsym}[1]{\ifmmode\begingroup\def\b@ld{bold}
  \text{\ifx\math@version\b@ld\bfseries\fi#1}\endgroup\else#1\fi}

\makeatother

\usepackage{babel}

\begin{document}

\title{Virtual Particle Interpretation of Quantum Mechanics - a non-dualistic
model of QM with a natural probability interpretation}

\author{J. M. Karimäki, Aalto University, Finland}

\date{7 June 2012}
\maketitle
\begin{abstract}
An interpretation of non-relativistic quantum mechanics is presented
in the spirit of Erwin Madelung's hydrodynamic formulation of QM 
and Louis de Broglie's and David Bohm's pilot wave models.
The aims of the approach are as follows: \emph{1) to have a clear
ontology for QM, 2) to describe QM in a causal way, 3) to get rid
of the wave-particle dualism in pilot wave theories, 4) to provide
a theoretical framework for describing creation and annihilation of
particles, and 5) to provide a possible connection between particle
QM and virtual particles in QFT.} These goals are achieved, if the
wave function is replaced by a fluid of so called \emph{virtual} particles.
It is also assumed that in this fluid of virtual particles exist a
few \emph{real} particles and that only these real particles can be
directly observed. This has relevance for the measurement problem
in QM and it is found that quantum probabilities arise in a very natural
way from the structure of the theory. The model presented here is
very similar to a recent computational model of quantum physics
and recent Bohmian models of QFT.\end{abstract}
\begin{description}
\item [{Keywords:}] Pilot wave, Madelung fluid, Bohmian mechanics,
Ontological interpretation of quantum mechanics, Hydrodynamic quantum mechanics
\end{description}

\section{Introduction}

Louis de Broglie \cite{key-21} and David Bohm \cite{key-2} have
shown that it is possible to give an interpretation of non-relativistic
quantum mechanics with a clearly defined ontology and continuous particle
trajectories. The models of de Broglie and Bohm, although developed
independently, are essentially the same, that is: there is a wave
function that evolves according to the Schrödinger equation and guides
the particles along continuous tracks. This theory, which we call
the pilot wave theory in this paper, produces the same predictions
as the standard interpretation of QM. The pilot wave theory has many
virtues, but it has also been criticized, among other things, for
having a kind of dualistic ontology, i.e.~needing both waves and
particles to describe the world. Also, it has been difficult to extend
the theory to describe the creation and annihilation of particles,
although recently some progress towards that goal has been made \cite{key-18,key-20}.

The interpretation, which we shall present in this paper removes the
problem of ontological dualism by removing the status of the wave
function as a fundamental building block of the theory. The wave function
is replaced by a fluid of particles, which can exist in two different
states: \emph{virtual} and \emph{real}. The number of real particles
is assumed to be finite, whereas the virtual particles form a continuum,
and thus the bulk of the fluid. All the particles in the fluid follow
the trajectories given by the pilot wave theory. The density of the
(mostly virtual) particles at some point gives the probability density
of finding a \emph{real} particle at that point at the time of measurement.

Since there is no wave function, the particles must somehow guide
themselves as a fluid. The equations for such a quantum fluid motion
were first introduced by Erwin Madelung in his hydrodynamic formulation
of quantum mechanics \cite{key-1}. The theory presented in this paper
can be understood as Madelung's theory supplemented by some minimal
additional definitions and assumptions needed to make it really work
-- in answering the measuring problem of QM, for example.

%
\begin{comment}
according to a criterion, which is local in the one particle case
and non-local in the N-particle case. Also the N-particle case can
be thought to be local in the configuration space. 
\end{comment}
{}

In the following sections we will see in more detail how the non-relativistic
single particle quantum mechanics is handled in this model. After
the detailed description of the single particle case, the generalization
to the $N$-particle case will be shortly explained.

%
\begin{comment}
The model presented here is very similar to a recent computational
model of quantum physics and chemistry \cite{key-22} and recent Bohmian
models of QFT \cite{key-18,key-20}.

The movement along Bohmian trajectories arises from a combination
of classical forces and a quantum force exerted by the continuum on
the real particles.

It may also help in filling the gap between particle QM and QFT by
giving rise to a class of particles bearing a close resemblance to
the virtual particles of QFT and also by raising hopes for a model
of particle creation and annihilation in particle QM.
\end{comment}
{}

\includegraphics[width=0.95\linewidth]{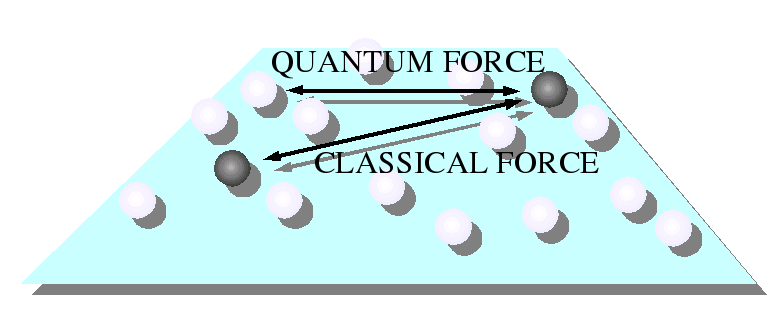}
\begin{description}
\item [{Figure}] 1. Schematic view of the interactions between {}``real''
particles (dark gray) and {}``virtual'' particles (light blue).
\end{description}
%
\begin{comment}
The empty and full particles could also be called \emph{virtual} and
\emph{real}, possibly a useful analogy when thinking about QFT. Other
possible names could be \emph{passive} and \emph{active}, or \emph{detectable}
and \emph{non-detectable}.
\end{comment}
{}

\section{Ontology of the model}

It is assumed that there is a background space, where the particles
move. The particles can have mass and electric charge. Spin is not
considered. The particles can be in two different states that we call
here virtual and real. In addition to the background space and particles,
there can also be external potential fields. These are assumed to
be essentially classical. The number $N$ of real particles in a given
volume is assumed to be finite. The number $N_{0}$ of virtual particles
in a given volume is expected to be much larger, essentially they
can be thought of as forming a continuum characterized by a fluid
density $\rho(\mathbf{x},t)$, and, consequently, a number density
$N_{0}\rho(\mathbf{x},t)$.

\section{Interactions}

Only the real particles can be observed directly. Otherwise all the
particles, virtual or real, follow continuous tracks that are the
same as those given by the de Broglie--Bohm pilot wave theory. Also
the velocities are assumed to be the same as in the pilot wave theory.
How this is possible without explicitly assuming the existence of
the wave function will be shown in the next section. It will also
be shown that the force affecting the particles can be understood
as the sum of classical forces caused by the real particles and quantum
forces caused by the fluid of virtual particles (Figure 1).

\section{Equations for single particle case}

The single particle case here means that there is only one real particle
in the system. The number of virtual particles can be assumed to be
practically infinite. The velocity of the particles, virtual or real,
is given by:

\begin{equation}
\mathbf{v}=\frac{1}{m}\nabla S,\label{velocity}\end{equation}
where $S(\mathbf{x},t)$ is simply the phase of the wave function
$\Psi$ multiplied by $\hbar$. It appears when the wave function
is written in the polar form: $\Psi=\rho^{1/2}e^{iS/\hbar}.$ There
are two equations that govern the evolution of the system:\begin{equation}
\frac{\partial\rho}{\partial t}=-\nabla\cdot(\frac{1}{m}\rho\nabla S)\label{conservation}\end{equation}

\begin{equation}
\frac{\partial S}{\partial t}=-\frac{(\nabla S)^{2}}{2m}-V+\frac{\hbar^{2}}{2m}\frac{\nabla^{2}\sqrt{\rho}}{\sqrt{\rho}}.\label{dynamical}\end{equation}

These are known as the hydrodynamic equations of QM, first introduced
by Erwin Madelung in 1926 \cite{key-1}, and later used by Louis de
Broglie \cite{key-21} and David Bohm \cite{key-2} in their pilot
wave models of QM. The hydrodynamic equations of QM are completely
equivalent to the Schrödinger equation given that some additional
constraints on the solutions $\rho(\mathbf{x},t)$ and $S(\mathbf{x},t)$
are imposed \cite{key-23}. The last term in (\ref{dynamical}), taken
with negative sign, is known as the quantum potential:\begin{equation}
Q=-\frac{\hbar^{2}}{2m}\frac{\nabla^{2}\sqrt{\rho}}{\sqrt{\rho}}.\label{quantum potential}\end{equation}

Now, we would like to give up using the phase $S$ (since it is derived
from the wave function), and, instead of it, take the velocity field
$\mathbf{v}(\mathbf{x},t)$ as the second variable. We replace $\nabla S$
by $m\mathbf{v}$ in (\ref{conservation}) and (\ref{dynamical})
and take the gradient of (\ref{dynamical}) to get the following two
equations:\begin{equation}
\frac{\partial\rho}{\partial t}=-\nabla\cdot(\rho\mathbf{v})\label{conservation b}\end{equation}
\begin{equation}
m\frac{\partial\mathbf{v}}{\partial t}=-\nabla(\frac{1}{2}m\mathbf{v}^{2}+V+Q).\label{velocity field}\end{equation}
Since we have taken a gradient, and want to ensure consistency with
the results of standard QM, we need the following additional constraint:
$\oint m\mathbf{v}\cdot d\mathbf{l}=2\pi n\hbar$, $n\in\mathbb{Z}$,
where the integration is along any closed smooth path along which
\textbf{$\mathbf{v}$} is well defined. Equation (\ref{conservation b})
can be interpreted as the conservation of probability, which is the
usual standpoint taken in hydrodynamic models of QM, or as the conservation
or matter, which is actually more suitable for the interpretation
presented in this paper. Equation (\ref{velocity field}) takes a
more familiar form, if the time derivative of \textbf{$\mathbf{v}$}
is taken along the particle trajectory:\begin{equation}
m\frac{d\mathbf{v}}{dt}=-\nabla(V+Q)\label{newton a}\end{equation}
or\begin{equation}
m\mathbf{a}=\mathbf{F}_{Classical}+\mathbf{F}_{Quantum},\label{newton b}\end{equation}
where $\mathbf{F}_{Classical}=-\nabla V$ is the classical force and
$\mathbf{F}_{Quantum}=-\nabla Q$ is a quantum force caused by the
local variations of $\rho$.

\section{Interpretation of quantum probability in the single particle case}

In QM it is customary to interpret the square of the absolute value
of the wave function $\rho=|\Psi(\mathbf{x},t)|\lyxmathsym{\texttwosuperior}$
to be the probability density of finding a particle at the location
$\mathbf{x}$, when its position is measured at the time $t$. This
law is held to be true in the standard interpretations of QM as well
as in the pilot wave approaches of de Broglie and Bohm, and it is
usually taken to be a basic fundamental feature of Quantum Mechanics.

In the standard interpretations, no further causal factors are sought
to explain this probability law, whereas in the pilot wave models
the probability law arises simply from our ignorance of the true particle
positions. Thus, in the pilot wave models, we have to \emph{assume}
that the probability density is equal to $|\Psi(\mathbf{x},t_{0})|\lyxmathsym{\texttwosuperior}$
initially. Once this assumption is made, the guidance condition ensures
that the probability density will agree with the value $|\Psi(\mathbf{x},t)|\lyxmathsym{\texttwosuperior}$
at any later time $t>t_{0}$.

In the model presented in this paper, the situation is, however, slightly
different. The probability density is again given by the density $\rho(\mathbf{x},t)$,
but it has a very natural explanation: Since both the real and virtual
particles follow the same equations of motion and their only difference
is that the real particle can be detected and the virtual particles
cannot, we simply assume that the single real particle can be any
one among all the particles forming the quantum fluid with number
density $N_{0}\rho(\mathbf{x},t)$. Thus, thinking by purely classical
definitions of probability, the fluid density $\rho(\mathbf{x},t)$
is the only natural candidate for the probability density of the real
particle. This situation is in no way different from finding one fish
that has swallowed a gold coin in a big swarm of fishes swimming in
the ocean and described by a {}``fish density'' $\rho_{\mathrm{fish}}(\mathbf{x},t)$,
or spotting one ringed bird flying in a big flock of other birds without
a ring.

\section{Generalization to $N$-particle case}

The generalization to the $N$-particle case, (where $N$ is the number
of real particles) simply requires the single particle wave function
to be replaced by the $N$-particle wave function. This wave function
can then be replaced by the $N$-particle density and velocity fields,
analogously to the single particle case. In this case, the single
particle is replaced by an $N$-particle multiplet: there will be
one real, detectable, $N$-multiplet moving in a continuous {}``fluid''
of virtual, undetectable $N$-multiplets. Since the fluid of $N$-multiplets
has dimension $3N$, the term \emph{hyperfluid} is here suggested to describe
it. (The name \emph {Madelung fluid} has been previously used,
but mainly in the context of superfluids in 3D or as an
abstract probability density fluid of arbitrary dimension.)

It can be shown that in some cases, the $N$-particle case reduces
to many separate one particle cases. Such is the case for $N$ free
particles. An analogous situation exists in superfluids. Actually,
a superfluid, such as He II, is a real world working model of the
virtual particle fluid presented in this paper.

\section{Particle creation and annihilation}

The second remaining problem of the pilot wave theory, \emph{vis \`{a}
vis} the difficulty of handling cases where particles are created
or destroyed, can also be solved, at least conceptually, using the
virtual particle interpretation. All we need is some kind of a process
of energy exchange between real and virtual particles. 

The following gives a qualitative picture of such a situation: Let
us assume, for example, that in the beginning there is a real electron--positron
pair, i.e.~two active particles and a sea of passive virtual particles,
of various kinds, all around them. The two active particles would
collide with sufficient energy and transfer their energy to a sea
of passive muon--anti-muon pairs. (This could exist without difficulty
since it would not be observable at low energies). Now, a passive
muon--anti-muon pair would receive the energy of the electrons and
become active, so that we would end up with a real muon and a real
anti-muon speeding in opposite directions from the point of impact.

However, at this point it must be emphasized that the process of energy
exchange described in the above example cannot be directly described
using the formalism presented in the earlier sections, but the model
would have to be extended or modified somehow (e.g., using the Fock
space). The reason for this is that particle creation and annihilation
cannot be described by non-relativistic particle QM. Instead one should
use a QFT version of the present theory, or some other clever formalism.
A Bohmian version of QFT with similarities to the model presented
here, and allowing particle creation and annihilation, has been described
in \cite{key-18}, so the possibility of devising such a formalism
also within the virtual particle scheme presented in this paper, doesn't
appear to be totally unrealistic. 

In case such a detailed model of particle creation and annihilation
can be produced, it could perhaps also be extended to model the Hawking
radiation close to a black hole. Related to this, we can also assume
that the virtual particles, or the quantum potential, have a small
rest mass. This might provide an answer to the mystery of dark mass
and dark energy in cosmology.

\section{Discussion}

The interpretation presented in this paper is in many ways similar
to other hydrodynamic or stochastic models \cite{key-1}-\cite{key-7}.
The differences in the approaches usually concern such things as the
role of the wave function, the nature of the particles (provided they
exist at all), and the form of the particle trajectories. Many approaches
assume very irregular particle trajectories or fluid motions to explain
the emergence of quantum probabilities \cite{key-9,key-4}. Why this
is not necessary was explained in section 5, where the quantum probability
law was found to be a natural consequence of the formalism. 

Madelung's original hydrodynamic formulation of QM \cite{key-1} can
be seen as a conceptual starting point for this work, and the two
approaches are quite similar%
\footnote{Among the more recent approaches, the closest in semblance to the
model presented in this paper is, quite surprisingly, a model introduced
by Courtney Lopreore and Robert Wyatt for numerical computations in
Quantum Physics and Chemistry \cite{key-22} (The existence of such
a method actually came as a big surprise to the writer). Comparing
an \emph{interpretation} of QM with a \emph{computational method}
is challenging, but at least the animations created by the computational
research group give an idea of the flow of a virtual particle fluid. %
} in spirit. Madelung's paper, however, leaves open many questions
relevant for the measurement problem in QM. These questions can be
answered within the framework of the proposed interpretation -- to
a large degree thanks to the features it inherits from the pilot wave
theory.

The virtual particle interpretation can be seen to be analogous to
the many worlds formalism. The difference is that the multitude of
worlds is here replaced by the multitude of the virtual particles
(or virtual $N$-particle systems). The fact that hydrodynamical equations
\cite{key-16} appear also in the decoherent histories approach may
have some significance. The situation can be expected to be similar
also in some of the spontaneous localization models.

Since the theory presented here is non-relativistic, it would be interesting
to study what the requirement of Lorentz invariance would bring into
the model. Another interesting topic would be to investigate the possibility
of formulating a Machian \cite{key-6} version of this theory and
include gravity in it. The connection of Feynman paths to the particle
paths in various hydrodynamical models is also worth investigating.

\section{Conclusions}

The model presented here, apart from the more speculative ideas of
sections 6 and 7, is a well-defined theory. The main motivation behind
it was to solve the problem of wave-particle dualism present in the
de Broglie--Bohm pilot wave theory, while retaining many of the desirable
features of that approach. One of the outcomes of this approach was
a surprisingly simple explanation for quantum probabilities. Since
the equations are the same as in the hydrodynamic formulation of QM
and the pilot wave theory, and the predictions the same as those given
by standard QM, the theory presented in this paper should be viewed
rather as a model, or, as the title suggests, as an interpretation
of quantum mechanics than as a new theory. However, by providing a
new perspective and giving new insights, this model can serve as a
basis for new theoretical developments.

%
\begin{comment}
filling gaps between old existing theories, such as QM and QFT, or
perhaps even giving answers to hitherto poorly understood or unexplained
phenomena.
\end{comment}
{}

\section{Acknowledgments}

The writer wishes to thank Lee Smolin, Hrvoje Nikoli\'{c}, Sam Karvonen
and Veikko Karimäki for useful ideas and comments.

\end{document}